\documentclass[prl,preprint,superscriptaddress,showpacs,floatfix]{revtex4}

\usepackage{amsmath}
\usepackage{graphicx}
\usepackage{dcolumn}
\usepackage{bm}
\usepackage{times}

\usepackage{anysize}
\usepackage{setspace}
\usepackage{framed}
\usepackage{fancyhdr}
\usepackage{longtable}

\RequirePackage{lineno}

\begin{document}



\title{Self-organization of network dynamics into local quantized states}

\author{Christos Nicolaides}
\affiliation{Sloan School of Management, Massachusetts Institute of Technology, Cambridge, MA, USA.}
\author{Ruben Juanes}
\affiliation{Department of Civil and Environmental Engineering, Massachusetts Institute of Technology, Cambridge, MA, USA.}
\author{Luis Cueto-Felgueroso}
\affiliation{Department of Hydraulics, Energy and Environment, Technical University of Madrid, Spain.}
\affiliation{Department of Civil and Environmental Engineering, Massachusetts Institute of Technology, Cambridge, MA, USA.}

\date{\today}

\begin{abstract}
Self-organization and pattern formation in network-organized systems emerges from the collective activation and interaction of 
many interconnected units. A striking feature of these non-equilibrium structures is that they are 
often localized and robust: only a small subset of the nodes, or cell assembly, is activated. Understanding the role 
of cell assemblies as basic functional units in neural networks and socio-technical systems emerges as a fundamental 
challenge in network theory. A key open question is how these elementary building blocks emerge, and how they operate, 
linking structure and function in complex networks. Here we show that a network analogue of the Swift-Hohenberg continuum 
model---a minimal-ingredients model of nodal activation and interaction within a complex network---is able to produce a complex 
suite of localized patterns. Hence, the spontaneous formation of robust operational cell assemblies in complex networks can be 
explained as the result of self-organization, even in the absence of synaptic reinforcements. Our results show that these self-organized, 
local structures can provide robust functional units to understand natural and socio-technical network-organized processes.
\end{abstract}


\maketitle

\section{Introduction}

Pattern formation in reaction-diffusion systems~\cite{Turing52,othmer_scriven_jtb_1971} has emerged 
as a mathematical paradigm to understand the connection between pattern and process in natural and sociotechnical systems~\cite{umulis_othmer_bmb_2015}.
The basic mechanisms of pattern formation by local self-activation and lateral inhibition, or short-range positive feedback and long-range negative 
feedback~\cite{gierer-meinhardt_kyb1972,meinhardt-gierer_bioessays2000} are ubiquitous in ecological and biological spatial systems, from morphogenesis and 
developmental biology~\cite{Turing52,umulis_othmer_dev_2013} to adaptive strategies in living organisms~\cite{kondo10,burger_pd2013} and spatial heterogeneity in predator-prey systems~\cite{Mimura78}. 
Heterogeneity and patchiness in vegetation dynamics, associated with Turing patterns in vegetation dyanmics have been proposed 
as a connection between pattern and process in ecosystems~\cite{klausmeier1999,hillerislambers2001}, suggesting a link between spatial vegetation patterns and 
vulnerability to catastrophic shifts in water-stressed ecosystems~\cite{vonhardenberg2001,rietkerk2004,kefi2007}.

The theory of non-equilibrium self-organization and Turing patterns has been recently extended to network-organized natural and socio-technical systems~\cite{moore_physd_2005,nakao10,hata_epl_2012,hata_scirep_2014}, 
including complex topological structures such as multiplex~\cite{asllani_pre_2014,kouvaris_scirep_2015}, directed~\cite{asllani_natcomm_2014} and cartesian product networks~\cite{asllani_scirep_2015}.
Self-organization is rapidly emerging as a central paradigm to understand neural computation~\cite{ermentrout_reprogphys_1998,garcia_fcn_2012,hutt_ptrsb_2014}. The dynamics of neuron activation, 
and the emergence of collective processing and activation in the brain, are often conceptualized as dynamical processes in network theory~\cite{bullmore09,chialvo10,bullmore12}. 
Self-organized activation has been shown to emerge spontaneously from the heterogenous interaction among neurons~\cite{hutt_ptrsb_2014}, and is often described as pattern 
formation in two-population networks~\cite{hansel-mato_prl_2001,blomquist_physd_2005,wyller_physd_2007,folias_prl_2011}. 
Localization of neural activation patterns is a conceptually challenging feature in neuroscience. Cell assemblies, or small subsets of 
neurons that fire synchronously, are the functional unit of the cerebral cortex in the Hebbian theory of mental representation and learning~\cite{palm81,kelso86,lansner02, buzsaki10,quiroga12}. 
Associative learning forms the basis of our current understanding of the structure and function of neural 
systems~\cite{reijmers07,neves08,lansner09}. It is also the modeling paradigm for information-processing artificial neural 
networks~\cite{clark_physmedbiol_1991,Izhikevich08,espinosa_prl_2015}. The emergence of cell assemblies in complex neural networks is a fascinating 
example of pattern formation arising from the collective dynamics of interconnected units~\cite{laing_neuralc_2001,hutt_ptrsb_2014}. Understanding the mechanisms 
leading to pattern localization remains a long-standing problem in neuroscience~\cite{kelso86,fuster00,laing_neuralc_2001,han07,silva09,gallistel13}. 

Here we show that simple mechanisms of nodal interaction in heterogeneous networks allow for the emergence of robust local activation patterns 
through self-organization. The simplicity and robustness of the proposed single-species pattern-forming mechanisms suggest that analogous dynamics 
may explain localized patterns of activity emerging in many network-organized natural and socio-technical systems. We demonstrate that robust local, 
quantized activation structures emerge in the dynamics of network-organized systems, even for relatively simple dynamics. 
We propose a minimal-ingredients, phenomenological model of nodal excitation and interaction within a network with heterogeneous connectivity. 
Our goal is to demonstrate that a simple combination of local excitation of individual units, combined with generic excitatory/inhibitory interactions 
between connected units, leads to self-organization, and can explain the spontaneous formation of cell assemblies without the need for 
synaptic plasticity or reinforcement. Our model can be understood as a network analogue of the Swift-Hohenberg continuum model~\cite{hilali_pre_1995,tlidi94,hilali_pre_1995,burke06}, 
and is able to produce a complex suite of localized patterns. The requirements are minimal and general: simple local dynamics based on 
canonical activation potentials, and interactions between nodes that induce short-range anti correlation and long-range correlation in activation. 
Because of their robustness and localization, self-organized structures may provide an encoding mechanism for information processing and 
computation in neural networks.

\section{Model of network dynamics and stability analysis}

\begin{figure}[http]
\centering
 \includegraphics[width=5.75in]{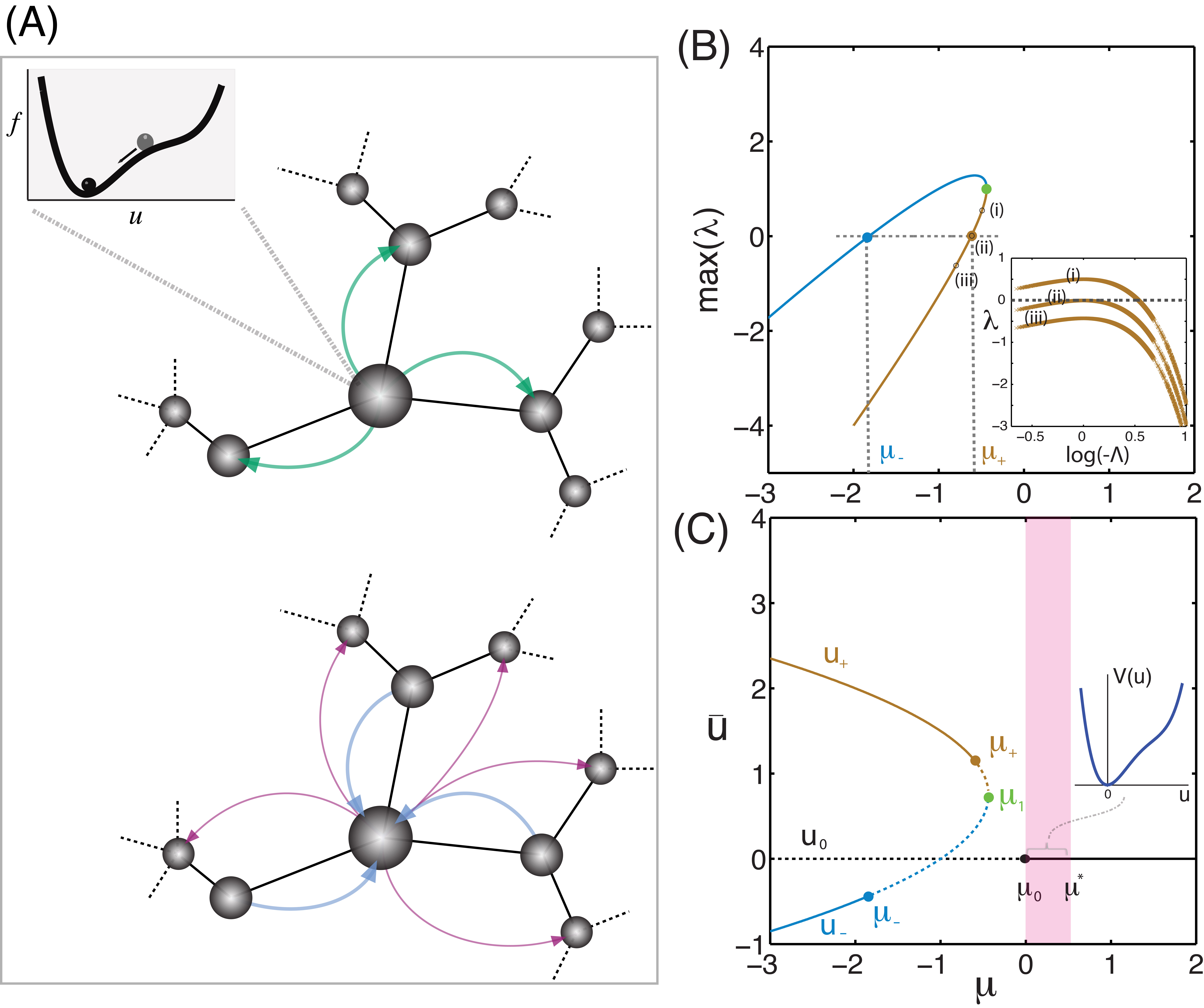}
  \caption{{\bf (A)} \emph{Pictorial illustration of our dynamical model of network interactions.} Locally, the nodal activation state is driven by the dynamic forcing term~$f(u,\mu)$. 
  In the inset we show the potential landscape---minus the integral of $f$ with respect to u, which exhibits a single well (at~$u=0$) with an inflection point, a necessary 
  condition for localized patterns to exist. Nodes interact in the network through diffusion-like exchanges via the links connecting them. The network Laplacian 
  operator,~$L^{(2)}$, represents short range diffusion of the species in the system (top). The network bi-Laplacian operator,~$L^{(4)}$, induces short range anti-correlation 
  with the nearest-neighbors, and long-range correlation with the second-nearest neighbors (bottom). {\bf (B--C)} \emph{ Linear stability analysis of the flat stationary 
  solutions of our model.} {\bf (B)} The maximum value of the growth rate~$\lambda$ as a function of the bifurcation parameter~$\mu$ for the two flat stationary 
  states~$u_{+}$ (brown) and~$u_{-}$ (blue) on a Barab{\'a}si-Albert network model with mean degree~$\langle k \rangle=3$ and size~$N$=2000. When the maximum 
  value of~$\lambda$ is negative, the state is stable with respect to small non uniform perturbation. (Inset) The growth rate~$\lambda$ as a function of the Laplacian 
  eigenvalue~$\Lambda$ (Eq. (5)) for three different values of the bifurcation parameter~$\mu$ as they indicated in the main diagram for the flat stationary 
  solution~$u_{-}$. {\bf (C)} The flat stationary solutions~$u_{0}$ and~$u_{\pm}$ as a function of~$\mu$ on the same network. Solid (dotted) lines represent stability (instability) 
  with respect to small non-uniform perturbations. The labelled bifurcation points are~$\mu_{0}=0$,~$\mu_{1}=-0.44$ and~$\mu_{+}=-0.62$ and~$\mu_-=-1.82$. The pink 
  shaded region is where we observe localized self-organization patterns with respect to the trivial solution~$u_{0}$. For values of~$\mu$ outside that region we get 
  either global activation patterns (for~$\mu<\mu_{0}$) or any perturbation relaxes back to the flat stationary solution (for~$\mu>\mu^{*}$).}\label{fig:fig1}
\end{figure}

We restrict our analysis to the simplified case of symmetric networks, but our main results can be generalized to other network topologies, including 
directed~\cite{asllani_natcomm_2014} and multiplex~\cite{asllani_pre_2014,kouvaris_scirep_2015} networks. 
A node's state of activation, measured through a potential-like variable~$u$, is driven by local excitation dynamics and by the interaction with other 
nodes in the network via exchanges through the links connecting them. In dimensionless quantities, the proposed excitation-inhibition model for the 
evolution of potential,~$u_{i}$, in each node~$ i=1,\dots,N$, is given by the 
model
\begin{equation}
\frac{du_{i}}{dt}=f(u_{i},\mu)+I_i,\label{eq:potential}
\end{equation}
where~$f(u_{i},\mu)$ is a dynamic forcing term, representing a double well potential, and~$\mu$ is a bifurcation parameter that will be used to establish the conditions for stability and localization of the response 
patterns~(Fig.~1A). The currents,~$I_{i}$, represent the excitatory/inhibitory interactions among nodes in the network. The structure of these nodal interactions 
is one of the key pattern forming mechanisms in the present model. We consider short-range anti-correlation, and higher-order, longer-range dissipative interactions. 
This two-level interaction structure, which induces anti correlation in the short range (nearest-neighbors, or first-order connectivity), and long-range correlation 
(second-nearest neighbors, or second-order connectivity) is represented in Fig.~1A. Mathematically, we express the integration of synaptic contributions as
\begin{eqnarray}\label{eq:interactions}
I_i= -2\sum_{j=1}^{N}L^{(2)}_{ij}u_{j}-\sum_{j=1}^{N}L^{(4)}_{ij}u_{j}.
\end{eqnarray}
The structure of the above nodal interactions turns the dynamics \eqref{eq:potential}--\eqref{eq:interactions} into a network anologue of the 
Swift-Hohenberg  continuum model, 
\begin{equation}
\frac{\partial{u}}{\partial{t}}=f(u,\mu)-(1+\nabla^2)^2 u,
\end{equation}
which is a paradigm for pattern-forming systems~\cite{hilali_pre_1995,tlidi94,burke06,lloyd10,gomez12}. 
The simplest form for the interaction matrices representing these correlation/anti-correlation effects (while ensuring that the interaction 
fluxes conserve mass or charge) is based on network representation of Laplacian and bi-Laplacian operators,~$\bm{L}^{(2)}$ and~$\bm{L}^{(4)}$, 
respectively. The network Laplacian,~$\bm{ L}^{(2)}$, is a real, symmetric and negative semi-definite~$N$$\times$$N$ matrix, whose elements are given 
by~$L_{ij}^{(2)}=A_{ij}-k_{i}\delta_{ij}$~\cite{nakao10}, where~$A_{ij}$ is the adjacency matrix of the network,~$k_{i}=\sum_{j=1}^{N}{A_{ij}}$ is the 
degree (connectivity) of node~$i$ and $\delta_{ij}$ is the Kronecker delta. A diffusive, Fickian-type flux of the activation potential~$u$ to node~$i$ is expressed 
as~$\sum_{j=1}^{N}{L_{ij}^{(2)}u_{j}}=\sum_{j=1}^{N}{A_{ij}(u_j-u_i)}$ (see Figure~\ref{fig:fig1}A--top). Plain waves and wavenumbers on a network topology 
are represented by the eigenvectors~$\bm{ \phi}^{\alpha}=(\phi_{1}^{(\alpha)},..,\phi_{N}^{(\alpha)})$ and the eigenvalues~$\Lambda_{\alpha}$ and of the Laplacian matrix, 
which are determined by the equation~$\sum_{j=1}^{N}L_{ij}^{(2)}\phi_{j}^{(\alpha)}=\Lambda_{\alpha}\phi_{i}^{(\alpha)}$, with~$\alpha=1,..,N$~\cite{othmer_scriven_jtb_1971}. 
All eigenvalues are real and non-positive and the eigenvectors are orthonormalized as~$\sum_{i=1}^{N}{\phi_{i}^{(\alpha)}\phi_{i}^{(\beta)}}=\delta_{\alpha,\beta}$, 
where~$\alpha,\beta=1,\dots,N$. The elements of the bi-Laplacian matrix of a network can be expressed as~$L^{(4)}_{ij}=\sum_{l=1}^N L_{il}^{(2)}L_{lj}^{(2)}=(A^2)_{ij}-(k_{i}+k_{j})A_{ij}+k_i^2\delta_{ij}$ where the $(A^2)_{ij}=\sum_{l}A_{il}A_{lj}$ 
matrix has information about second order nodal connectivity and takes nonzero values if 
node~$i$ is two jumps away from node~$j$. The operation~$\sum_{j=1}^{N}L_{ij}^{(4)}u_{j}$ models negative diffusion (inhibition) from the first neighbors of 
node~$i$ and at the same time diffusion from its two-jump neighborhood (see Figure~\ref{fig:fig1}A--bottom). The bi-Laplacian,~$\bm{L}^{(4)}$, has the same 
eigenvectors as~$\bm{L}^{(2)}$ (i.e. ${\bm{\phi}}^{\alpha}$) and its eigenvalues are the square of those of~$\bm{L}^{(2)}$, $\Lambda_{\alpha}^2$.

To understand the properties and pattern-forming mechanisms in our model, we first investigate the stability of flat states of the dynamical system \eqref{eq:interactions}:
\begin{equation}
\frac{du_{i}}{dt}=f(u_{i},\mu)-2\sum_{j=1}^{N}L^{(2)}_{ij}u_{j}-\sum_{j=1}^{N}L^{(4)}_{ij}u_{j},\quad\quad i=1,...,N.
\end{equation}
Flat, stationary solutions~$\bar{u}$ of Eq.~(4) satisfy~$f(u_{i},\mu)=0$, where the nodal state of activation is equal for all nodes in the 
network,~$u_{i}=\bar{u},\forall i=1,...,N$.  For~$f(u_i,\mu)=-(1+\mu)u_i+1.5u_{i}^2-u_{i}^3$, there are three uniform 
solution branches given by~$u_0=0$ and~$u_{\pm}=[1.5\pm\sqrt{1.5^2-4(\mu+1)}]/2$. It is well known in one and two dimensional 
continuum spaces that these uniform states can become unstable and a wealth of self-organized patterns can arise~\cite{tlidi94,hilali_pre_1995,burke06,lloyd10,gomez12}. 
In a linear stability analysis, the stability of flat stationary solutions to small perturbations is determined by the eigenvalues of the Laplacian 
and bi-Laplacian matrices. Introducing small perturbations,~$\delta u_{i} $, to the uniform state~$\bar{u}$, $u_{i}=\bar{u}+\delta u_{i}$, the linearized 
version of Eq.~(4) takes the form~$d\delta u_{i}/dt=f_{u}(\bar{u})\delta u_{i}-2\sum_{j=1}^{N}L^{(2)}_{ij}\delta u_{j}-\sum_{j=1}^{N}L^{(4)}_{ij}\delta u_{j}$, where~$f_{u}=\partial_{u}f=\frac{\partial f(u,\mu)}{\partial u}$.  
After expanding the perturbation~$\delta u_{i}$ over the set of the Laplacian eigenvectors,~$\delta u_{i}=\sum_{\alpha=1}^N q_{\alpha}e^{\lambda_{\alpha}t}\phi_{i}^{(\alpha)}$, where $q_{\alpha}$ is the expansion coefficient, 
the linearized equation is transformed into a set of~$N$ independent linear equations for the different normal modes:
\begin{equation}
\lambda_{\alpha}=f_{u}(\bar{u})-2\Lambda_{\alpha}-\Lambda_{\alpha}^2,\quad \alpha=1,\dots,N,
\end{equation}
where~$\Lambda_{\alpha}$ are the eigenvalues of the Laplacian matrix. 
The~$\alpha$-mode is unstable when Re~$\lambda_{\alpha}$ is positive. Instability occurs when one of the modes (the critical mode) begins to grow. 
At the instability threshold, Re~$\lambda_{\alpha}=0$ for some~$\alpha_{c}$ and Re~$\lambda<0$ for all other modes. In Figure~1B and C we summarize 
the linear stability analysis of the flat states of our model on a scale-free network constructed using the Barab{\'a}si-Albert model (BA) of network 
growth and preferential attachment~\cite{barabasialbert99}. We find that, indeed, there is a large parameter range for which the resting potential is stable. 
As we demonstrate below, in the stable regime, input stimuli may trigger localized patterns of activation.

\section{Localized patterns}

\begin{figure}[http]
\centering
\includegraphics[width=6.25in]{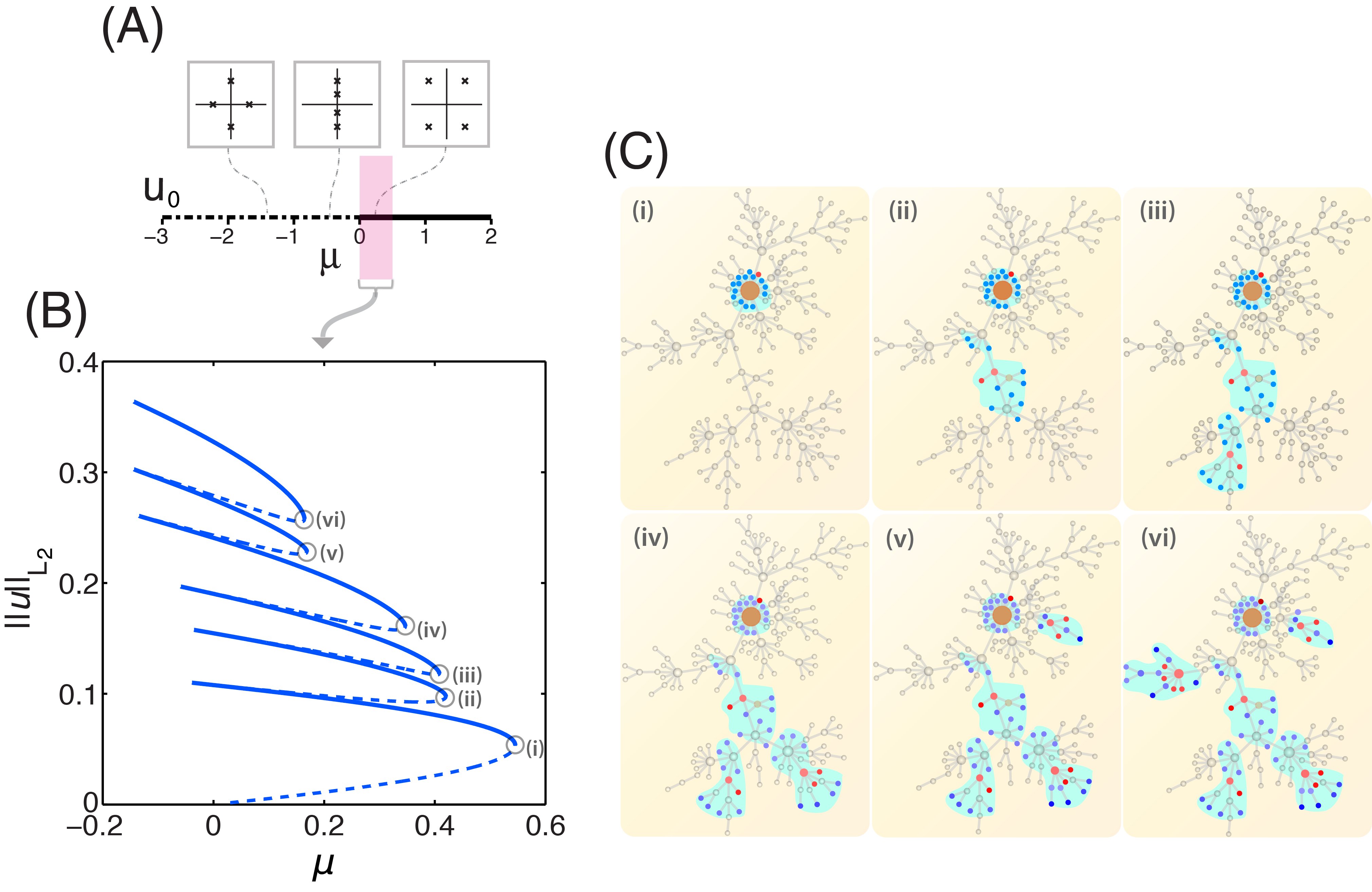} 
  \caption{\emph{Localized self-organized quantized  patterns.} {\bf (A)} Stability of the trivial flat stationary state of our model with respect to the values 
  of the bifurcation parameter,~$\mu$. For positive values of~$\mu$ the trivial stationary solution is stable with respect to uniform small random perturbations 
 (solid line) while for negative values of~$\mu$ this state becomes unstable (dotted line). Also shown in the insets are the topological eigenvalues of the 
  trivial state as we tune the bifurcation parameter. The behavior of eigenvalues in the neighborhood of~$\mu=0$ indicates the possibility for localized 
  patterns in the neighborhood of small positive values of~$\mu$ (pink shaded region). {\bf (B)} A single branch of the bifurcation diagram in a Barab{\'a}si-Albert 
  network model of size $N$=200 with mean degree equal to~$\langle k \rangle= 3$ and minimum degree equal to~1. Solid (dotted) lines represent stable (unstable) 
  localized solutions. {\bf (C)} Visualization of the localized patterns corresponding to the states indicated on the bifurcation diagram (B). Gray-colored nodes are 
  non-active ($u=0$), red-colored nodes are active with~$u>0$ and blue-colored nodes are active with~$u<0$. The size of the node is proportional to its eigenvalue 
  centrality.}\label{fig:fig2}
\end{figure}

Localized activation patterns are possible due to the particular structure of the model, with short- and long-range 
nodal interactions. Mathematically, the localized states are homoclinic orbits in the network space around the base resting state,~$\bar{u}=u_{0}=0$. 
The existence of these homoclinic orbits can be studied using the technology developed for the linear stability analysis.
Since homoclinic orbits leave the flat state as we approach a small neighborhood (cluster) of the network, the fixed point must have both stable 
and unstable eigenvalues. We linearize Eq.~(4) around~$u=0$ and write~$\delta u_i=\sum_{\alpha=1}^N q_{\alpha}\phi_{i}^{(\alpha)}$,~$q_{\alpha}\ll 1$, arriving 
to the relation~$f_{u}(0)-2\Lambda_{\alpha}-\Lambda_{\alpha}^2=0$. Since the Laplacian eigenvalues~$\Lambda_{\alpha}$ are real and non-positive values, 
we can write them in the form~$\Lambda_{\alpha}=-k_{\alpha}^2$. If~$\mu> 0$ the topological eigenvalues of~$u=0$ form a complex 
quartet~$k_{\alpha}=\pm i \pm \frac{\sqrt{\mu}}{2}+\mathcal{O}(\mu)$. For~$\mu=0$ they collide pairwise on the imaginary axis, and for~$\mu<0$ they 
split and remain on the imaginary axis~$k_{\alpha}= i (\pm1\pm \frac{\sqrt{-\mu}}{2})+\mathcal{O}(\mu)$. For~$\mu=-1$ two of the topological eigenvalues 
collide at the origin and for~$\mu<-1$ they move onto the real axis. These results are summarized in Figure~\ref{fig:fig2}A. The topological 
eigenvalues in the neighborhood of~$\mu=0$ are characteristics of the reversible~$1:1$ resonance bifurcation. Theory shows that under certain conditions 
the hyperbolic regime contains a large variety of topologically localized states~\cite{burke06}.

To understand the onset of localized patterns for different model parameters and input stimuli, we construct the bifurcation diagram of the resting state, 
as a function of the total potential energy of the stimulus and bifurcation parameter~$\mu$, in the vicinity of~$\mu\simeq 0$~\cite{burke06,lloyd10}. 
A single bifurcation branch---constructed using a pseudo-arclength continuation method~\cite{keller77}---has a characteristic ``snaking" 
structure of localized states with varying activation energy~$||{\bf u}^{0}||_{L2}=(1/N\sum_{j=1}^{N}{ u^{0}_j})^{1/2}$ (Fig.~\ref{fig:fig2}B). As the system 
jumps from one steady state branch to the next one, a new neighborhood in the network is being activated. Figure~\ref{fig:fig2}C 
visualizes the different steady localized states of the six different branches as they are spotted in the diagram of Figure~\ref{fig:fig2}B.
The response of the system is \emph{quantized}: the transition from one pattern of activation to another one is discontinuous as we vary the 
activation energy~$||{\bf u}^{0}||$, or the parameter~$\mu$ (Fig.~\ref{fig:fig2}B). These jumps in activation energy correspond to the addition of 
neighbor nodes to the cluster (Fig.~\ref{fig:fig2}C).

\begin{figure}[http]
\centering
 \includegraphics[width=5.25in]{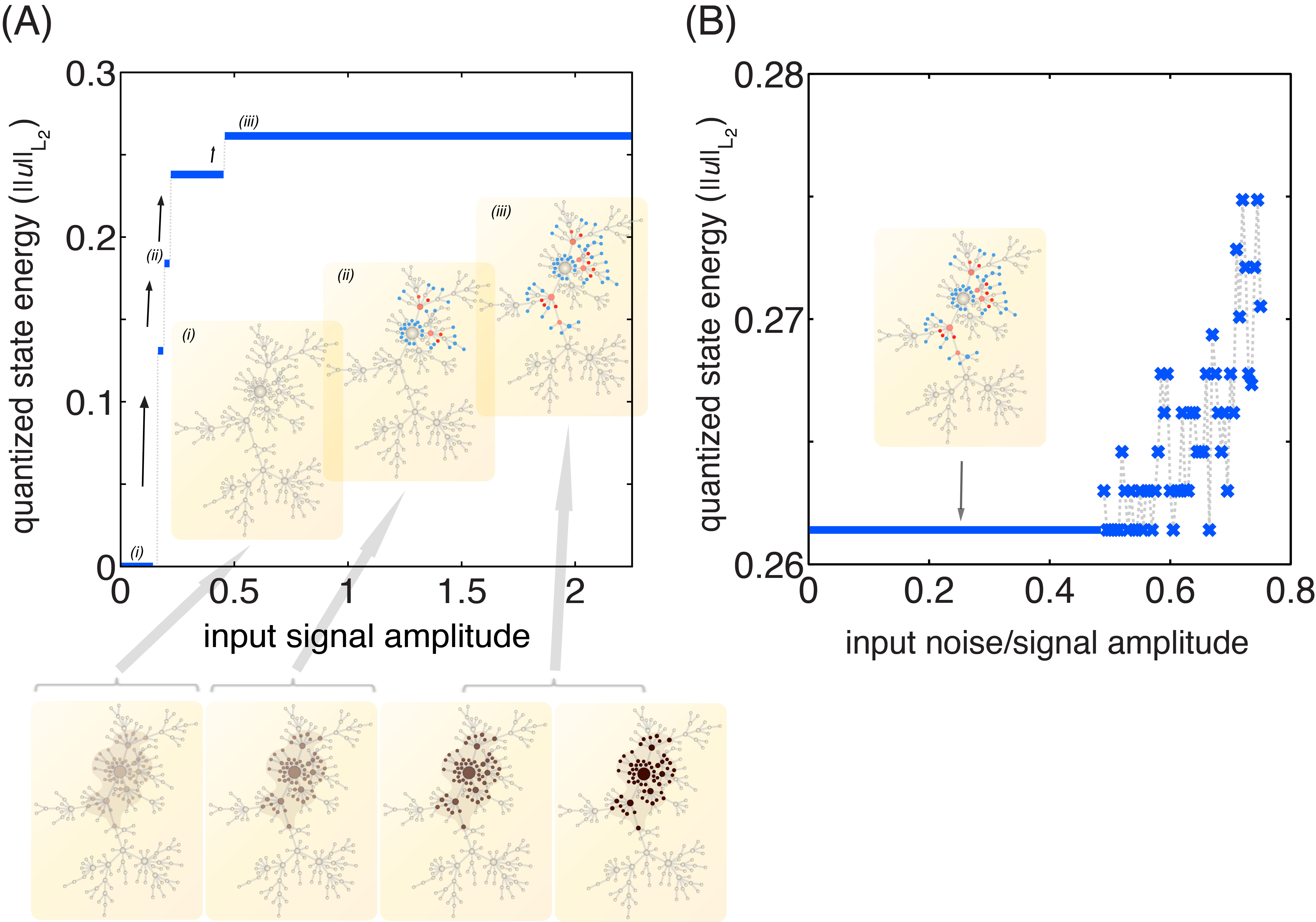} 
   \caption{\emph{Robustness of quantized patterns.} {\bf (A)} Energy of the resulting quantized state with respect to the input signal amplitude~$\hat{u}$ at the 
   nearest and next-nearest neighbors of the best connected node in the system. When the amplitude is very small, the initial perturbation relaxes back to the trivial 
   solution and no quantized state is formed (\textit{i}). As the amplitude of the input signal is increased, fragile quantized states are formed (\textit{ii}). When the 
   amplitude of the input signal is larger than a threshold value, a very robust quantized state is formed (\textit{iii}). Further increases in the input signal amplitude 
   lead to the same quantized state. (insets) Visualization of the input signal in our network topology (the amplitude increases from left to right) as well as the 
   resulting equilibrium state. {\bf (B)} The energy of the resulting quantized state with respect to the ratio between the signal amplitude and the noise amplitude. 
   Starting from the step-like input signal that gives the robust quantized state, we add random noise at the already perturbed neighborhood and we compute the 
   energy of the resulting quantized state over 100 realizations. We use a Barab{\'a}si-Albert scale-free network of size~$N$=200 and mean degree~$\langle k \rangle$=4.  
}\label{fig:fig3}
\end{figure}

The discontinuous---quantized---nature of the network response leads to \emph{robustness} in the local, final equilibrium patterns with respect 
to the input signal amplitude. To gain insight into the robustness of the localized patterns of activation, we performed a synthetic test in which we initially 
stimulate a specific neighborhood in the network, where we set~$u_i=\hat{u}\geq0$ (i.e. a step-like function signal in network topology) and let the system 
evolve to equilibrium without decay. We gradually increase the amplitude~$\hat{u}$ of the initial signal, and record the final energy values of the equilibrium, 
localized states. For small amplitudes the perturbation relaxes back to the resting state, and no activation pattern is elicited. There is a threshold in the 
energy of the input stimulus beyond which robust quantized states are form. The states are robust in the sense that further increments in the input signal 
amplitude do not change the final equilibrium pattern (Fig.~\ref{fig:fig3}A).

The self-organized local structures are also robust with respect to random noise in the initial stimulus. We perform Monte Carlo simulations that probe the 
impact of the noise-to-signal ratio on the energy of the emerging quantized state. We have confirmed that the presence of small-amplitude noise has no effect 
on the equilibrium states of nodal activity. As can be expected, we do observe a departure from the energy of the base equilibrium state when the 
noise-to-signal ratio is sufficiently large, thereby masking the base stimulus altogether (Fig.~\ref{fig:fig3}B).

\section{Mean-field approximation of the global activation patterns}

Our model predicts a range of parameter values where localized states disappear, and are replaced by \emph{global activation patterns}. 
Mathematically, global patterns are possible when the non-active stationary solution is perturbed outside the parameter region of localized 
patterns ($\mu<0$). These---global---Turing patterns~\cite{nakao10,Turing52} can be understood and modeled using the Mean-Field Approximation (MFA), a method 
that segregates nodes according to their degree and has been successfully used to approximate a wide variety of dynamical processes 
in heterogeneous networks, like epidemic spreading~\cite{pastorvespignani15,Nicolaides13,Nicolaides12}, activator-inhibitor models~\cite{nakao10} 
and voter models~\cite{baronchelli11}.

This theory allow us to reduce the problem to a single equation for the membrane potential for all the nodes in the system. Since in our model both the 
degree and two-jump degree play important role in the formation of patterns, we use a MFA where 
we assume that all the nodes with the same degree and two-jump degree behave in the same way. We start by writing Eq.~(3) in the form 
\begin{equation}
\frac{d u_{i}}{dt}=f(u_{i})-2(h_{i}-k_{i}u_{i})-(l_{i}-g_{i}-k_{i}h_{i}+k_{i}^{2}u_{i}),
\end{equation}
where the local fields felt by each node,~$h_{i}=\sum_{j=1}^{N} A_{ij}u_{j}$,~$l_{i}=\sum_{j=1}^{N} (A^2)_{ij}u_{j}$ and~$g_{i}=\sum_{j=1}^{N} A_{ij} (k_{j}u_{j})$ 
are introduced. These local fields are then approximated as~$h_{i}\simeq k_{i} H_{u}$,~$l_{i}\simeq k_{i}^{(2)}H_{uu}$ and~$g_{i}\simeq k^{(2)} H_{u}$, 
where~$k_{i}=\sum_{j=1}^{N}A_{ij}$ is the degree and~$k_{i}^{(2)}=\sum_{j=1}^{N}(A^2)_{ij}$ is the number of secondary connections of 
node~$i$ (two-jump degree). The global mean fields are defined by~$H_{u}=(1/N)\sum_{k}N_{k}H_{u}^{(k)}$ where~$H_{u}^{(k)}=(1/(kN_{k}))\sum_{i\epsilon k}\sum_{j} A_{ij}u_{j}$ 
and~$H_{uu}=(1/N)\sum_{k^{(2)}}N_{k^{(2)}}H_{uu}^{(k^{(2)})}$, where~$H_{uu}^{(k^{(2)})}=(1/(k^{(2)}N_{k^{(2)}}))\sum_{i\epsilon k^{(2)}}\sum_{j} (A^2)_{ij}u_{j}$. 
Here,~$N_{k}$ is the number of nodes with degree~$k$,~$N_{k^{(2)}}$ is the number of nodes with~$k^{(2)}$ number of two-jump neighbors 
and~$N=\sum_{k}N_{k}=\sum_{k^{(2)}}N_{k^{(2)}}$ is the size of the network. In the above expressions, with~$\sum_{i\epsilon k}$ we denote the sum over the 
nodes with degree~$k$ and with~$\sum_{i\epsilon k^{(2)}}$ the sum over the nodes with two-jump nodal connectivity~$k^2$.

With this approximation, the individual model equation on each node interacts only with the global mean fields~$H_{u}$ and~$H_{uu}$ and its 
dynamics is described by:
\begin{eqnarray}
 \frac{d u}{dt} = &f(u)&-2\alpha(H_{u}-u)-\\ \nonumber
&-&[\beta H_{uu}  -\alpha^2 H_{u}-\beta H_{u}+\alpha^2 u].
\end{eqnarray}
Since all nodes obey the same equation, we have dropped the index~$i$ and introduced the parameters~$\alpha(i)=k_{i}$ and~$\beta(i)=k_{i}^{(2)}$. The activation potential depends now on the global fields~$H_{u}$ and~$H_{uu}$ as well as on the parameter compination ($\alpha,\beta$), i.e. $u=u(H_{u},H_{uu},\alpha,\beta)$.
If the global mean 
fields~$H_{u}$ and~$H_{uu}$  are given, the combination ($\alpha,\beta$) plays the role of a bifurcation parameter that controls the dynamics of each node in the system. The time independent version of above 
mean field equation can be written as a third degree algebraic equation that we solve~$N$ times for the~$N$ nodes in the system. For each node~$i$, we get three solutions~$u_i^{l}$, $l=1..3$ 
that can be stable or unstable depending on the sign (negative or positive respectively) of the operator~$f'|_{u_i^{l}}+2\alpha-\alpha^2$. 

After tuning the bifurcation parameter~$\mu$ to a negative value, we can compute the global Turing pattern from direct numerical simulations and determined 
the global mean fields~$H_{u}$ and~$H_{uu}$. Each node~$i$ in the network is characterized by its degree and second nodal connectivity, so that it possesses 
a certain parameter combination, ($\alpha,\beta$). Substituting these computed global mean fields as well as the values of~$\alpha$ and~$\beta$ into equation (7), 
bifurcation diagrams of a single node can be obtained and projected onto the Turing pattern. In Figure~(4A) we show for our ``toy network model" that the stable 
brunches of the nodal bifurcation diagrams calculated using the MFA fit very well the computed Turing pattern. We further assess the dependence of the network topology on the global pattern formation and we find that when the degree distribution is narrower compared to a scale-free network, the distribution of the ($\alpha,\beta$) is more homogeneous and therefore the stationary Turing patterns look smoother (Figure~4B and C).  Therefore, global network Turing patterns are 
essentially explained by the bifurcation diagrams of individual nodes coupled to the global mean fields, with the coupling strength determined by their degree 
and two-jump connectivity. 

\section{Conclusions}

Our results suggest a new mechanism for the formation of localized nodal assemblies in networks, arising from long-range---second neighbor---interactions. 
Rather than relying on reinforcing mechanisms---synaptic plasticity, we show that localized, robust nodal assemblies are possible due to self-organization. 
The emergence of localized activation patterns derived from the simple and general functional structure of our proposed conceptual model: local dynamics 
based on activation potentials, and interactions between nodes that induce short-range anticorrelation and long-range correlation in node-to-node exchanges. 
The proposed system is a network analogue of the Swift-Hohenberg continuum model, and is able to produce a complex suite of robust, localized patterns. 
Hence, the spontaneous formation of robust operational cell assemblies in complex networks can be explained as the result of self-organization, even in the 
absence of synaptic reinforcements. Hence, these self-organized, local structures can provide robust functional units to understand natural 
and technical network-organized processes.

\begin{acknowledgments}
Funding for this work was provided by an MIT Vergottis Graduate Fellowship and a McDonnell Postdoctoral Fellowship in Studying Complex Systems (to CN), the US Department of Energy through a DOE CAREER Award (grant DE-SC0003907) and a DOE Mathematical Multifaceted Integrated Capability Center (grant DE-SC0009286) (to RJ), and a Ram\'on y Cajal Fellowship from the Spanish Ministry of Economy and Competitiveness (to LCF).
\end{acknowledgments}

\section{Author contribution statement}
CN, RJ and LCF designed the research, performed the analysis and wrote the manuscript.

\section{Additional Information}
The authors declare no competing financial interests.

\pagebreak

\begin{figure}[http]
\centering
 \includegraphics[width=4.75in]{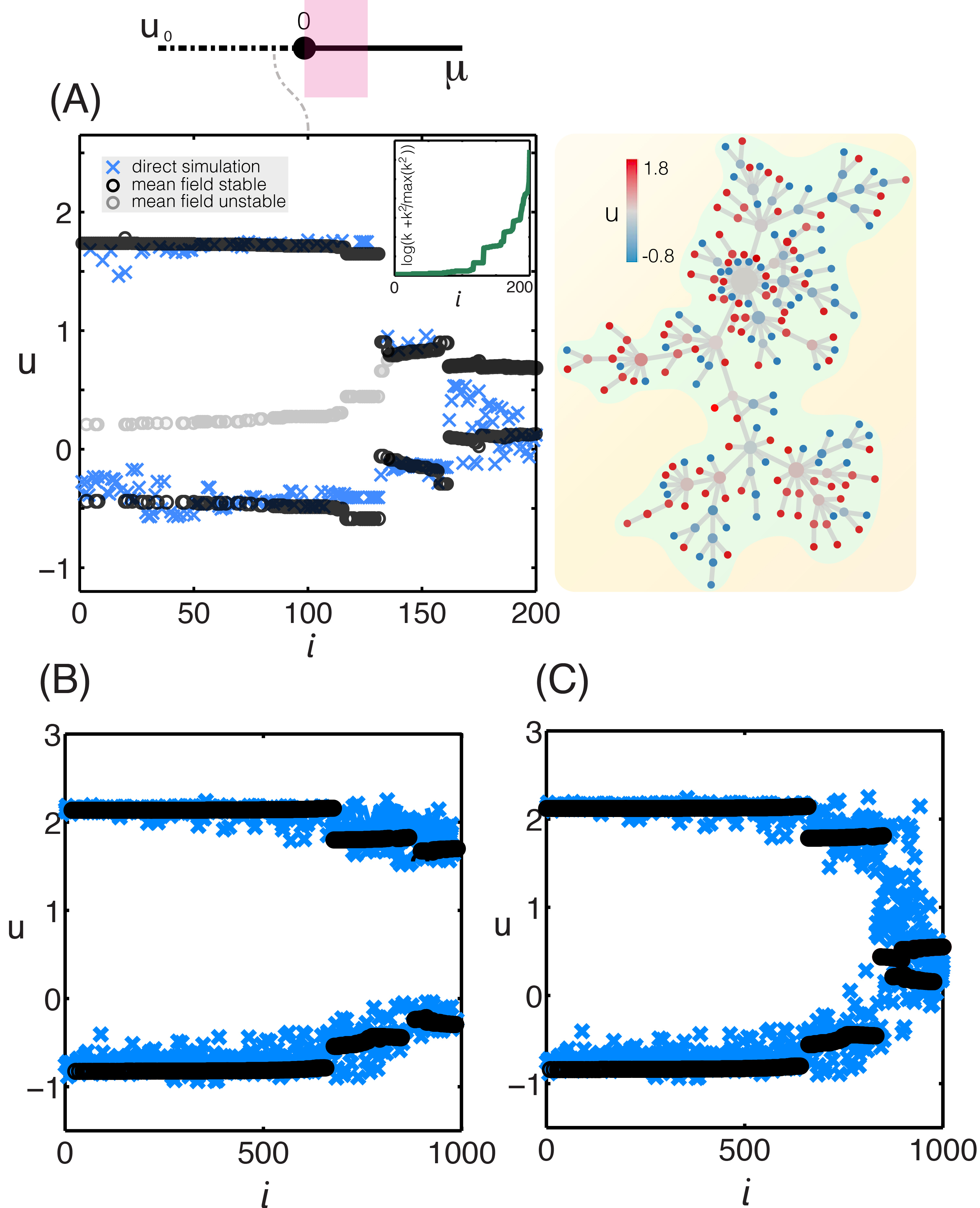} 
  \caption{\emph{Global Turing patterns.}  Global patterns are possible when the non-active stationary solution is perturbed outside the parameter region of 
  localized patterns ($\mu<0$). The initial exponential growth of the perturbation is followed by a nonlinear process leading to the formation of stationary Turing 
  patterns. {\bf (A)} (left) The activation profile as a function of the node index~$i$ of a global stationary Turing pattern from direct simulation (blue crosses) is compared 
  with the mean-field bifurcation diagram. Black curves indicate stable branches while grey curves correspond to unstable branches of a single activator--inhibitor 
  system coupled to the computed global mean fields. We sort the node index~$i$ in increasing connectivity~$k$. Nodes with the same degree are sorted with 
  increasing two-jump connectivity~$k^{(2)}$ (see Inset). We use the same Barab{\'a}si-Albert network model as in Fig.~2 and we set the bifurcation parameter 
  equal to~$\mu=-1/4$. We have confirmed that similar results hold for larger network sizes. (right) Visualization of the global activity pattern on the network
  topology. {\bf (B)} The activation profile as a function of the node index $i$ of global stationary Turing patterns from direct simulation for bifurcation parameter $\mu=-0.25$ on an  Erd{\"o}s-R{\'e}nyi random network with size $N=1000$ and mean degree $\langle k \rangle=4$ (blue curve) along with the stable branch of the mean field approximation (black curve). {\bf (C)}  The activation profile as a function of the node index $i$ of global stationary Turing patterns from direct simulation on a Barab{\'a}si-Albert scale free network with the same mean degree and the same number of node as the Erd{\"o}s-R{\'e}nyi of B.  We sort the node index in increasing connectivity $k$ and two-jump connectivity $k^{(2)}$. }\label{fig:fig4}
\end{figure}

\end{document}